\documentclass{ws-p8-50x6-00}

\newcommand{\Z      }[4]{\ensuremath{#1\,\pm #2\,^{+\,#3}_{-\,#4}}}
\newcommand{\epem   }{\ensuremath{\mathrm{e}^+\mathrm{e}^-}}
\newcommand{\aem    }{\ensuremath{\alpha}}
\newcommand{\invpb  }{\ensuremath{\mathrm{pb}^{-1}}}
\newcommand{\qsq    }{\ensuremath{Q^{2}}}
\newcommand{\qzm    }{\ensuremath{\langle \qsq \rangle}}
\newcommand{\ft     }{\ensuremath{F_{2}^{\gamma}}}
\newcommand{\ftxq   }{\ensuremath{\ft(x,\qsq)}}
\newcommand{\ftn    }{\ensuremath{F_{2}^{\gamma}/\aem}}
\newcommand{\ftqn   }{\ensuremath{\ft(\qsq)/\aem}}
\newcommand{\gevsq  }{\ensuremath{\mathrm{GeV}^2}}
\newcommand{\kt     }{\ensuremath{k_{\mathrm{t}}}}

\newcommand{\xvis   }{\ensuremath{x_{\mathrm{vis}}}}
\newcommand{\fitres }{\ensuremath{\ftqn=(\Z{0.049}{0.021}{0.049}{0.037})%
           +(\Z{0.139}{0.007}{0.009}{0.013})\,\ln\,\qsq}}
\def\rightnote{UCL/HEP 2001-03: {\it To appear in the proceedings of Photon 2001.}}

\begin{document}

\title{Measurement of the Hadronic Photon Structure Function \ft\ at LEP2}

\author{R. J. Taylor}

\address{Department of Physics and Astronomy, University College London,\\
Gower Street, London WC1E 6BT, United Kingdom\\
E-mail: rjt@hep.ucl.ac.uk}


\maketitle

\abstracts{
 The hadronic structure function of the photon \ftxq\ is measured 
 as a function of Bjorken $x$ and of the factorisation scale \qsq\
 using data taken by the OPAL detector at LEP. 
 Previous OPAL measurements of the $x$ dependence of \ft\ are extended to an 
 average \qsq\ of 767~\gevsq.
 The \qsq\ evolution of \ft\ is studied for $11.9 < \qzm < 1051$~\gevsq.
 As predicted by QCD, the data show positive scaling violations in \ft.
 Several parameterisations of \ft\ are in agreement with the measurements
 whereas the quark-parton model prediction fails to describe the data.}
\section{Introduction}
 Much of the present knowledge of the structure of the photon has been
 obtained from measurements of deep-inelastic electron-photon scattering at 
 \epem\ colliders.
 With the high statistics and high electron energies at LEP2 it is possible
 to study \ft\ at $\qsq > 1000$~\gevsq\ and to determine the evolution of 
 \ft\ with the factorisation scale.
 \par
 The determination of \ft\ uses the fact that the differential cross-section
 of the $e\gamma$ DIS reaction as a function of \qsq\ and Bjorken $x$
 is proportional to \ftxq~\cite{NIS-9904}.
 For finite \qsq\ the absolute normalisation of \ft\ cannot be predicted 
 by perturbative QCD and has to be determined from data, but its 
 evolution with \qsq\ is predicted by QCD to be 
 logarithmic.
 \par
 This analysis~\cite{PN489} is based on 632~\invpb\ of data  taken by 
 the OPAL detector in the years 1997--2000, with \epem\ 
 centre-of-mass energies ranging between 183 and 209~GeV.
 It extends the measurements of \ft\ as a function of $x$ 
 up to $\qzm=767$~\gevsq, and significantly improves on the 
 precision of the measurement of the \qsq\ evolution of \ft.
%
%
\section{Data selection}
 Three samples of events are studied in this analysis, classified
 according to the subdetector in which the scattered electron is observed. 
 Electrons are tagged using the SW (33-55~mrad), FD (60-120~mrad) and EE 
 (230-500~mrad) subdetectors.
 Events are selected by applying cuts on the energy and polar angle of the 
 scattered electrons and on the invariant mass and multiplicity of the 
 hadronic final state. 
 An anti-tag condition is applied to ensure that the virtuality of the 
 quasi-real photon is small.
 In addition, for the EE sample an isolation criterion is applied to the
 tagged electron.
 \par
 The number of events passing the cuts is 27819, 11874 and 414 for the
 SW, FD and EE samples respectively. The data range in \qsq\ from 
 7.1--2323~\gevsq.
%
%
\section{Results}
 The analysis presented here addresses two questions:
 first, the extension of the measurement of \ft\ as a function of $x$ to 
 the highest possible value of \qzm\ using the EE sample; and second, the
 evolution of \ft\ with \qsq\ at medium values of $x$ based on all three 
 samples.
 \par
 Based on the \xvis\ distribution in each range of \qsq, the structure function
 \ft\ has been obtained from the data by unfolding.
 No attempt has been made in this analysis to access the region of $x<0.1$, so
 using a one dimensional unfolding on a linear scale in $x$ is appropriate,
 in contrast with the previous OPAL analysis of \ft~\cite{OPALPR314}. 
 For this purpose the RUN program~\cite{BLO-8401BLO-9601} has been used. 
 To obtain the central values the HERWIG 5.9+\kt(dyn)~\cite{MAR-9201}
 program was used as the input Monte Carlo model to the unfolding.
 \par
\begin{figure}[htb]
\begin{center}
\epsfxsize=22pc 
\epsfbox{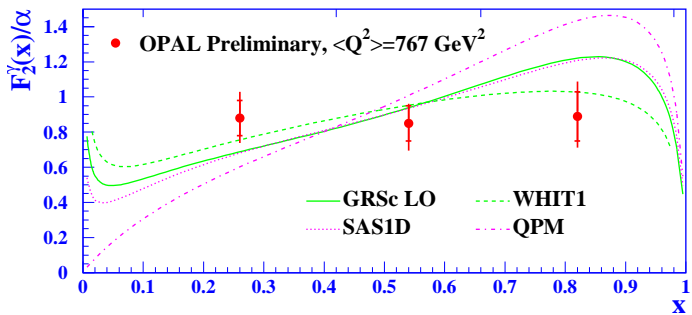} 
\end{center}
\caption{The measured \ftn\ as a function of $x$ at $\qzm=767$~\gevsq.
         The inner error bars indicate the statistical error and the full
         bars the total error.
         The tick marks at the top of the figure represent the bin boundaries.}
\label{fig:pn489_04}
\end{figure}

 After subtraction of background, the EE sample has been unfolded on a 
 linear scale in $x$ using three bins in $x$ spanning the range 
 $0.1-0.98$. Each data point is corrected for radiative effects using the 
 RADEG program~\cite{LAE-9603LAE-9701} and bin-centre corrections are applied.
 The result for \ftn\ is shown in Figure~\ref{fig:pn489_04}.
 \par
\begin{figure}[htb]
\begin{center}
\epsfxsize=24pc 
\epsfbox{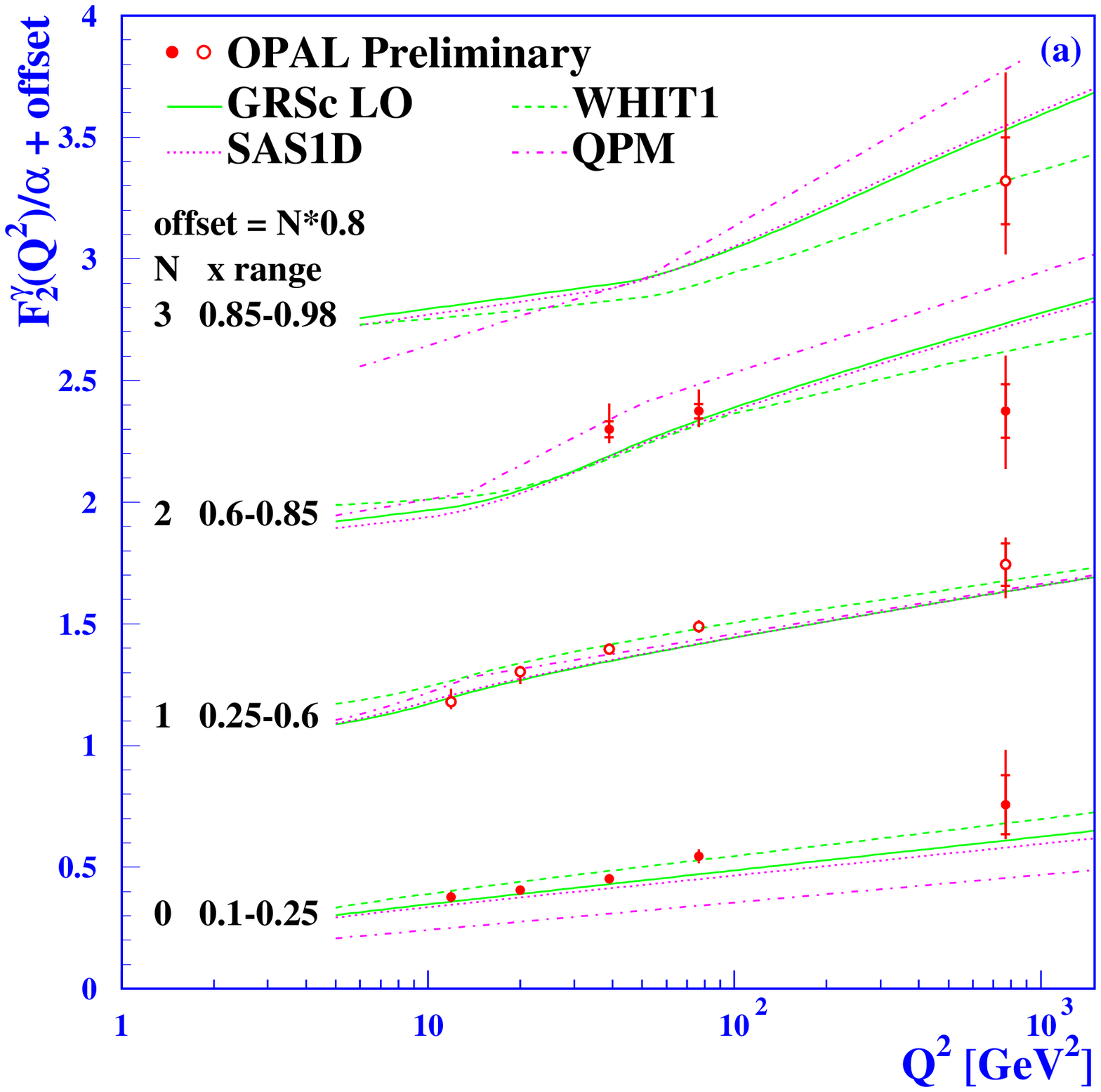} 
\epsfxsize=24pc 
\epsfbox{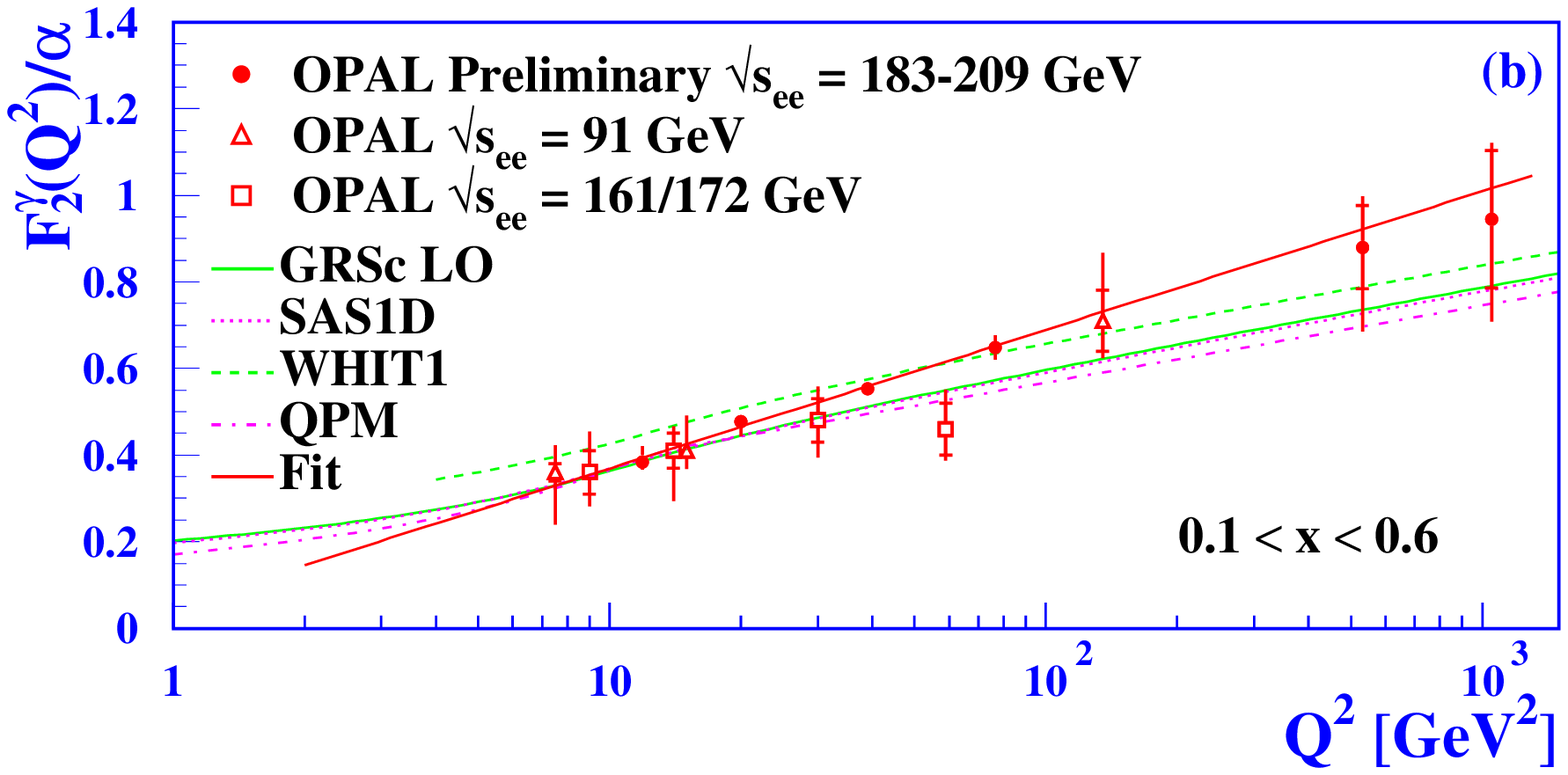} 
\end{center}
\caption{The evolution of \ftn\ as a function of \qsq\ for several bins
         of $x$, (a) 0.10--0.25, 0.25--0.60, 0.60--0.85 and
         0.85--0.98 and (b) for the central region 0.10--0.60.
         The inner error bars indicate the statistical error and the full
         bars the total error.}
 \label{fig:pn489_05}
\end{figure}

 The evolution of \ft\ 
 with \qsq\ has been measured for several $x$ ranges using all three samples.
 Due to the large statistics the SW and FD samples are further split into 
 two bins of \qsq.
 The data are unfolded separately in each bin of \qsq\ and
 corrected for radiative effects.
 The results are shown in Figure~\ref{fig:pn489_05}.
 The data in Figure~\ref{fig:pn489_05}(a) show positive scaling violations
 in \ft\ for the $x$ ranges 0.10--0.25 and 0.25--0.60, as predicted by QCD.
 For the range 0.60--0.85, within statistics, the data are compatible 
 with scaling violations.
 \par
 To quantify the slope for medium values of $x$, where data is available
 at all values of \qsq, a linear function of the form 
 $a + b\,\ln\,\qsq$ has been fitted to the data in the region 
 0.10--0.60, Figure~\ref{fig:pn489_05}(b).
 For this investigation the EE sample is also divided into two regions 
 in \qsq.
 The result of the fit is $$\fitres\,,$$
 where \qsq\ is in \gevsq.
 This is in agreement with the the previous OPAL value~\cite{OPALPR207},
 and the errors on $a$ and $b$ have been strongly reduced.
 \par
 Both for the measurement of \ft\ at $\qzm=767$~\gevsq\ and for the 
 investigation of the \qsq\ evolution of \ft, the quark-parton model 
 prediction is not in agreement with the data.
 It shows a much steeper rise  than the data as a function of $x$ 
 for $\qzm=767$~\gevsq\ and also a different behaviour in the \qsq\ evolution.
 In contrast, the GRSc~\cite{GLU-9902}, SaS1D~\cite{SCH-9501} and 
 WHIT1~\cite{HAG-9501} parameterisations of \ft\ are 
 much closer to the data, with the WHIT1 prediction giving the best 
 description of the data.
 This means that the corresponding parton distribution functions of the 
 photon are adequate at large values of $x$ and at factorisation scales
 up to about 1000~\gevsq.
%
%

\end{document}